# Superconductivity in Layered van der Waals Hydrogenated Germanene at High Pressure

*Yilian Xi[1,3†], Xiaoling Jing[2†], Zhongfei Xu[1,3,4†], Nana Liu[3], Yani Liu[1,3], Miao-Ling Lin[5], Ming Yang[1], Ying Sun[1], Jincheng Zhuang[1,3], Xun Xu[1,3], Weichang Hao[1,3], Yanchun Li[6], Xiaodong Li[6], Ping-Heng Tan[5], Quanjun Li[2\*], Bingbing Liu[2\*], Shi Xue Dou[1,3], Yi Du[1,3\*]*

[1] School of Physics and BUAA-UOW Joint Research Centre, Beihang University, Beijing 100191, China
[2] State Key Laboratory of Superhard Materials, Jilin University, Changchun 130012, China
[3] Institute for Superconducting and Electronic Materials (ISEM), Australian Institute for Innovative Materials (AIIM), University of Wollongong, Wollongong, NSW 2500, Australia
[4] Beijing Computational Science Research Centre (CSRC), Beijing 100193, China
[5] State Key Laboratory of Superlattices and Microstructures, Institute of Semiconductors, Chinese Academy of Sciences, Beijing 100083, China
[6] Beijing Synchrotron Radiation Facility, Institute of High Energy Physics, Chinese Academy of Sciences, Beijing 100049, China

\* To whom correspondence should be addressed: yi_du@uow.edu.au (Y.D.), liubb@jlu.edu.cn (B.L.) and liquanjun@jlu.edu.cn (Q.J.L.)
† These authors contributed equally to this work.

Keywords: superconductivity, germanene, hydride, two dimensional, van der Waals, high pressure

Structural and superconducting transitions of layered van der Waals (vdW) hydrogenated germanene (GeH) were observed under high-pressure compression and decompression processes. GeH possesses a superconducting transition at critical temperature ($T_c$) of 5.41 K at 8.39 GPa. A crystalline to amorphous transition occurs at 16.80 GPa while superconductivity remains. An abnormally increased $T_c$ up to 6.1 K has been observed in the decompression process while the GeH remained amorphous. Thorough *in-situ* high-pressure synchrotron X-ray diffraction and *in-situ* high-pressure Raman spectroscopy with the density functional theory simulations suggest that the superconductivity of GeH should be attributed to the increased density of states at the Fermi level as well as the enhanced electron-phonon coupling effect under high pressure. The decompression-driven superconductivity enhancement arises from pressure-induced phonon softening related to an in-plane Ge-Ge phonon mode. As an amorphous metal hydride superconductor, GeH provides a platform to study amorphous hydride superconductivity in layered vdW materials.

**1. Introduction**

Germanene, a germanium-based layered material in a low-buckled honeycomb lattice, possesses Dirac electronic states in the low-energy region.[1-8] Exotic properties, including topological insulating states and the quantum spin Hall effect, are expected to emerge in this two-dimensional (2D) material, which has attracted tremendous attention due to its rich physics and promising potential applications.[9,10] Its mixed $sp^2$/$sp^3$ hybridization state, however, may lead to a structural instability that prevents exploration of its novel properties. Recently, hydrogenated germanene (GeH) has been successful synthesized, in which the hydrogen atoms terminate on top of the Ge sites in germanene and form an ordered hexagonal sublattice.[11-15] The hydrogen stabilizes the 2D structure of germanene by donating electrons to saturate the dangling bonds of the surface. GeH is a fully hydrogenated germanene and the repeat unit contains two layers in the $c$ direction. In this system, individual Ge atom in each layer is directly bonded with three germanium atoms in the $a$-$b$ plane and with a hydrogen atom in the $c$ axis. In this case, the hydrogen atoms are arranged alternately on both sides of the layer to saturate germanene, resulting in the opening of a direct band gap at the Γ point in the Brillouin zone (BZ). It makes GeH a promising candidate for electronic and optoelectronic applications such as field effect transistors,[16,17] lithium ion batteries,[13] and photodetectors.[18]

Theoretical studies have proposed that superconductivity may be induced in 2D Dirac materials with a honeycomb lattice through chemical decoration if the adatoms form an ordered structure.[19-23] Even though GeH meets such criteria, its semiconducting nature leads to zero density of states (DOS) at the Fermi level ($E_F$), and no superconductivity has ever been observed yet. High pressure, as an important thermodynamic parameter, has been emerging as a feasible tool to explore extraordinary properties by modulating the crystal and electronic structures of materials. Pressure-induced superconducting transitions have been confirmed in various elements.[24-27] For example, the insulators at ambient pressure exhibit superconductivity when sufficient compression is applied. Those studies have led to the discovery of high-temperature superconductivity in hydrides, *e.g.*, sulphur hydrides[28-30] and lanthanum hydrides.[31-34] It is well accepted that the superconducting transition can be given rise to an increased DOS at the $E_F$ enhanced electron-phonon coupling (EPC), and high phonon frequencies under high pressure. In terms of GeH, its EPC is extremely sensitive to strain and external pressure, which is attributed to its layered low-buckled structures similar to those in germanene. External pressure may also result in a reformation in the stacking order and interlayer twisting between neighbouring GeH layers, which, in turn, leads to enhanced out-of-plane phonon frequencies similar to those in the other 2D materials.[35] These features indicate that 2D GeH may undergo a superconducting transition under external pressure.

Here, we report pressure-induced superconductivity in 2D GeH single crystal. The pressure-driven phase transition from crystalline to amorphous occurs in compression at the superconducting critical temperature ($T_c$) of 5.4 K. An abnormally increased $T_c$ as high as 6.1 K is observed in the decompression process, in which the sample retains its 2D amorphous structure. Theoretical simulation indicates that the superconducting transition as well as the varied $T_c$ in compressed GeH are mainly due to the enhancement of the DOS at the $E_F$ and the EPC under high pressure, while the superconductivity of the amorphous GeH is likely due to the softened Ge-Ge mode. This work provides a platform for studying amorphous hydride superconducting phenomena.

## 2. Results and Discussion

GeH single crystal samples a few millimetres in size were prepared by the topochemical deintercalation method (**Figure 1b** inset). As illustrated in **Figure 1a**, GeH exhibits a typical 2D layered structure with vdW interaction between the H-Ge-H layers.[11-13] The low-buckled Ge plane in each layer is saturated by surface H atoms. **Figure 1b** shows the X-ray diffraction (XRD) pattern of the as-prepared GeH single crystal, in which all peaks can be indexed to a 2H phase in the space group P63/mc. Electron microscopy results, including transmission electron microscopy (TEM) and scanning electron microscopy (SEM) images, of cleaved GeH crystal demonstrate its 2D layered structure (**Figure 1d** and **1e**). The high-resolution TEM image of GeH in **Figure 1f** shows clear lattice fringes with the interplanar spacing of 0.339 nm, which corresponds to the (100) planes. This agrees well with the XRD and the selected area electron diffraction (SAED) results. The Raman spectrum of the GeH sample in **Figure 1c** shows two phonon peaks at 301.27 cm$^{-1}$ and 224.85 cm$^{-1}$, which are assigned to in-plane $E_{2g}$ and the out-of-plane $A_{1g}$ vibrational modes, respectively.[11] The X-ray photoelectron spectroscopy (XPS) spectrum of the sample shows two peaks at binding energy of 31.52 eV and 29.62 eV (**Figure 1g**), which are attributed to the Ge $3d_{5/2}$ and $3d_{3/2}$ orbitals, respectively. The Ge $3d_{5/2}$ and $3d_{3/2}$ peaks of GeH possess higher binding energy in comparison to pure germanium due to the Ge-H bonds. (For more sample information, see **Figure S1** to **S3**).

The temperature dependence of the electrical resistance $R(T)$ for a GeH single crystal under various pressures up to 34.11 GPa and released to 5.81 GPa was measured by using a diamond anvil cell (DAC, **Figure S4**). Electric resistance was measured using an inverting *dc* current in a van der Pauw technique implemented in a customary cryogenic setup (**Figure S5**). As shown in **Figure 2**, a superconducting transition emerges under the compression pressure of 8.39 GPa, which is demonstrated by a sharp drop of resistance to zero at $T_c$ of 5.41 K (as shown in the inset in **Figure 2b**). With the pressure gradually increased to 34.11 GPa, the superconductivity remains, but with a decreased $T_c$. **Figure S6** illustrates the temperature dependence of the electrical resistance in temperature range from room temperature down to 2.5 K. Surprisingly, $T_c$ increases in the decompression process, and reaches 6.1 K when the pressure was released to 5.81 GPa, as shown in **Figure 2a**. The $T_c$ is plotted as a function of pressure in **Figure 2b**, in which the $T_c$ shows two "linearly" decreasing regions over the entire compression region. It first decreased from 5.41 K to 3.61 K when the pressure was increased to 16.80 GPa, and further decreased to 3.31 K for the pressure up to 34.11 GPa. The pressure dependence shows that $dT_c/dP$ = -0.218 K/GPa and $dT_c/dP$ = -0.016 K/GPa below and above 16.80 GPa, respectively. Similar behaviour is shown in the temperature-dependence of the normalized electrical resistance (with resistances normalized by their values at $T$ = 300 K), as shown in **Figure S7**. To further illustrate the change in the superconducting state at about 16.80 GPa, the pressure dependence of the superconducting transition width ($\Delta T_c$), and the slope of $\Delta T_c$ vs pressure are also plotted in **Figure 2c**, in which a knee point at 16.80 GPa is demonstrated. High-pressure magnetoresistance measurement in external magnetic fields has been carried out to reveal superconducting behaviours in GeH under external pressure, as shown in **Figure S8**. It shows that $T_c$ is decreased with increased magnetic field. The upper critical magnetic field $H_{c2}(T)$ can be approximated by the relationship of $H_{c2}(T) = H_{c2}^* \times (1-T/T_c)^2$, which allows us to estimate the $H_{c2}$ is about 23.0 T at 14.62 GPa. It is worth to note that the upper critical fields obtained at 14.62 GPa is greater than the corresponding Pauli paramagnetic limits (~6.8 T). We summarize our experimental results on the pressure dependence of the obtained $T_c$ onset for the studied GeH in **Figure 2d**. To distinguish the germanium superconductivity, we also show the distict behaviours of superconducting transition temperature as a function of pressure of the GeH, amorphous germanium and single crystal germanium (**Figure S9**). Because GeH is a typical semiconductor,[18] the pressure-induced transition from a semiconducting state to conducting (metallic) state is of fundamental interest. We measured the pressure-dependent resistance of GeH sample at room temperature. As shown in **Figure S10**, the resistance decreased strikingly in the range of 6~8 GPa, which reflects that the semiconducting-to-metallic-state transistion of GeH occurs at 6 GPa and completes at 8 GPa. It is also supported by *R-T* and *I-V* results measured in high pressure.

In order to understand the nature of pressure-induced superconductivity in GeH, *in-situ* measurements with variable pressure, including high pressure synchrotron XRD (with incident wavelength $\lambda = 0.6199$ Å) and Raman spectroscopy characterisations were carried out on a tiny GeH single crystal at room temperature, as shown in **Figure 3**. The Rietveld refinement of the diffraction pattern at 8.39 GPa is shown as a typical example (**Figure S11**). Due to extremely small scattering factor of H atoms, they have almost no effect on the XRD patterns. We did not include H atoms in the refined structure model. Only the position changes of Ge atoms were considered. The structural evolution of GeH crystal in both the compression and the decompression processes is plotted in **Figure 3a-3c.** (The original 2D diffraction patterns and the corresponding integrated 1D diffraction pattern under different pressures are respectively provided in **Figure S12** and **S13**). Two regions can be identified in the pressure-dependent XRD results, which are a crystalline region (denoted as Phase I) and an amorphous region (denoted as Amorphous) with the boundary at 16.80 GPa. As the pressure increases, all diffraction peaks shift to a higher angle, indicating that the compression leads to a decrease in the corresponding *d*-spacing value, as displayed in **Figure 3a** and **3b**. According to the calculations, the *d*-spacing values of the (101) and (110) peaks respectively decrease by 0.2 Å and 0.1 Å during compression from 1 atm to 16.80 GPa. The initial 2H crystal structure of the GeH sample remains stable up to 16.80 GPa. Once the pressure rises above 16.80 GPa, significant broadening of the diffraction peak at 11º is observed, while all the other diffraction peaks disappear. This indicates that a possible phase transition, that is, pressure-induced amorphization, occurs in the GeH sample. Upon further increase of the pressure to 33.71 GPa, there is no obvious change in the XRD pattern, except that the intensity of the peak become slightly weaker, confirming that the GeH amorphous phase is stable over the investigated pressure range. The pressure-induced amorphization of GeH at high pressure is most likely attributable to its 2D layered structure with weak interlayer interaction. The possible interlayer stacking and rotation are also expected to vary during the compression. In the decompression process, the GeH sample remains amorphous. The broad peak around 11° splits into two peaks, which suggests that the amorphization transition is an irreversible process. It should be noted that the critical pressure point of 16.80 GPa agrees well with the feature in the $R(T)$ results, which suggests that the structural evolution and phase transition determine the superconductivity of GeH. **Figure 3d** shows the *in-situ* high-pressure Raman spectra. The Raman peaks are fitted with Lorentzian line-shapes. With increasing pressure, the $E_{2g}$ vibrational mode shows a shift from 301.0 cm$^{-1}$ at 0.27 GPa to 322.9 cm$^{-1}$ at 14.66 GPa. A Raman peak (marked with blue asterisk) has been observed which attributes to GeH layer sliding under high pressure, similar to that observed in molybdenum disulfide.[35] The Raman peak disappears at pressure of about 17 GPa, indicating the amorphization. With the pressure released, a soft phonon mode appears between 200 cm$^{-1}$ and 300 cm$^{-1}$. These Raman results are consistent with the high-pressure XRD results. **Figure 3e** shows the pressure dependence of the Raman peak positions for the GeH sample. The blue shift of the $E_{2g}$ mode arises from pressure-induced Ge-Ge bond contraction. A jump in peak position occurs at around 8.06 GPa, the point at which superconductivity appears in the GeH sample. The full width at half maximum (FWHM) values of the Raman peaks also exhibit a turning point at 8.06 GPa, as shown in **Figure 3f**. This suggests a possibly enhanced EPC effect in GeH due to the external pressure that is applied. It is known that several superconductors, *e.g.* FeSe, undergo a phase transition at low temperature with compression pressure.[27] In order to verify whether GeH also undergoes a structural phase transition upon cooling down at high pressure, an *in-situ* high-pressure Raman characterization over GeH sample has been carried out at 25 K. The results show similar feature to high-pressure Raman spectra measured at 300 K (**Figure S14**), which reflects that the GeH does not undergo additional phase transition except amorphization under compression pressure during cooling process. The stress gradients in compressing process is particularly important in high-pressure experiments, since they may lead to anomalies in the experimental data that can be incorrectly interpreted as new physical phenomena.[36, 37] We have carefully studied and ruled out impacts of stress gradient in our measurement (Supporting Information and **Figure S15**).

We performed DFT calculations to explore the origins of the pressure-induced superconductivity in GeH, as shown in **Figure 4. Figure 4a** and **4c** shows the crystal and electronic structures of GeH at ambient pressure and at 10 GPa, respectively. At ambient pressure, GeH is a semiconductor with a direct band gap of 0.83 eV, which agrees well with previous studies.[11,12] In the high pressure phase, GeH undergoes a dehydrogenation process with interlayer Ge-Ge binding and isolated $H_2$ dimers observed along the *c*-axis, which results in the metallic P63/mmc phase. Possible crystal structures and formation enthalpies of $Ge_xH_y$ were given in **Figure S16,** indicating that our predicted GeH structure is the most energy favourable under the pressure of 10 GPa. The DFT results were supported by *in-situ* Raman spectra (**Figure S14**), in which a vibrational mode of $H_2$ dimers have been identified in a range of 4200 $cm^{-1}$ to 4300 $cm^{-1}$ at the pressure greater than 7.7 GPa. This particular structural feature suggests that the compressed GeH may display similar physical properties to Ge, such as superconductivity. Two crossed bands at the $E_F$, mainly originating from Ge-*p* and H-*s* states, play crucial role in the increased electronic DOS (0.0612 states/eV/unit cell), which is believed to facilitate the superconducting transition. The total density of states and projected density of states are respectively given in **Figure S17** and **S18**, H-*s* strongly hybridized with Ge-$p_z$ at 0 GPa and transforms to slightly hybridize with Ge-$p_x$ and Ge-$p_y$ at 10 GPa, which indicate the Ge-H bond broken at high pressure. Energy of H-*s* electrons significantly downshift from deep conduction band to the $E_F$ confirms the arising contribution from H atoms in GeH at high pressure. We also performed phonon calculations in the thermodynamic stability range of GeH. As shown in **Figure 4b** and **4d**, the absence of any imaginary frequency modes in the BZ indicates dynamical stability in these structures. The partial phonon density of states reveals that the heavy Ge atoms dominate the low-frequency modes, while the light H atoms contribute significantly to the high-frequency modes. More dispersed bands in the BZ indicate the bulk structural properties for the high pressure phase. The disappearance of the frequency gap from 100 cm$^{-1}$ to 200 cm$^{-1}$ and the appearance of a frequency gap from 500 cm$^{-1}$ to 600 cm$^{-1}$ may be attributed to the formation of H$_2$ dimers, which will soften the modes of H phonons. In comparison, the *c*-axis mode A$_{1g}$ obviously stiffened because Ge-Ge bonds in the high pressure phase are much stronger than Ge-H bonds in the ambient pressure one, even though more soft modes are found in the low-frequency region in the high pressure GeH phase, which is regarded as an evidence of strong EPC. The similar structural features to Ge, flat energy bands near $E_F$, and significant soft phonon modes of GeH at high pressure may relate to its well-established superconducting behaviours. To better compare these results with experiments, the EPC parameter $\lambda$, the logarithmic average phonon frequency ($\omega_{\log}$), and the Eliashberg phonon spectral function $\alpha^2F(\omega)$[38] were investigated at 10 GPa. The resulting value of $\lambda$ is 0.707, which indicates that the EPC effect is fairly strong. It is known that $T_c$ can be estimated from the Allen-Dynes modified McMillan equation $T_c = \frac{\omega_{\log}}{1.2}\exp\left[\frac{1.04(1+\lambda)}{\lambda - u^*(1-0.62\lambda)}\right]$.[39] In this work, the caluclated $T_c$ is 7.7 K by using a typical Coulomb pseudopotential value of $u = 0.1$, which is in a good agreement with the experimental results.

Although amorphous hydride superconductivity has been reported, the previous examples all used the hydrogen-induced amorphization method to introduce hydrogen atoms into the crystal to form an alloy, [40,41] such as Pt-Si hydrides. In these hydrides, the introduction of hydrogen generally reduces $T_c$, and superconductivity is lost when the pressure is released. Moreover, it must involve a hydrogenation or hydrogen absorption process over these alloys at high pressure. This means these alloys were actually *in-situ* synthesized at high pressure and then showed superconductivity. They do not naturally existed and are not stable at ambient pressure. In contrast, GeH is a stable 2D vdW compounds at ambient pressure and demonstrates superconducting transition at high pressure without any *in-situ* hydrogenation/absorption process. Alternatively partial $H_2$ might escape from the edges. The formation of $H_2$ has been verified by high-pressure Raman spectra, in which the peak at ~4200 cm$^{-1}$ under 7.7 GPa (**Figure S14 a-c**) is assigned to vibrational mode of $H_2$. Surprisingly, the amorphous GeH sample retained its metallic conduction behaviour in the decompression process. The $T_c$ shows an abnormal increase of up to 6.10 K when the pressure is released down to 5.81 GPa, rather than showing reversible behaviour. We note that similar enhanced superconductivity behaviour has also been observed in various systems including elements,[24,25] copper oxides,[42] and various compounds such as $CaC_6$,[43] $In_2Se_3$,[44] and P:$BaNi_2As_2$[45]. In Bardeen-Cooper-Schrieffer (BCS) theory, the EPC parameter $\lambda$ scale is inversely proportional to the average squared phonon frequency.[46] In our case, the pressure-induced phonon stiffening accounts for the drop of $T_c$ above 16.80 GPa for the amorphous GeH. Conversely, the obverse $T_c$ enhancement in the decompression process is consistent with the Ge-Ge mode softening, further supporting the BCS mechanism of phonon mediated pairing. In addition, enhanced superconductivity caused by a disorder-induced multifractal wavefunction was found in monolayer niobium dichalcogenides.[47] The disorder level may also plays a subtle role in modulating the superconductivity behaviour under high pressure, which needs further study.

## 3. Conclusion

In conclusion, pressure-induced superconductivity in 2D GeH is reported, accompanied by a structural phase transition from a crystalline phase to an amorphous phase near the pressure of 16.80 GPa. The onset of superconductivity in GeH occurs at 8.39 GPa with a $T_c$ of 5.41 K. The superconductivity arises from the increasing DOS at the $E_F$ and the enhanced EPC effect in the compression process, which is confirmed by both *in-situ* experimental and theoretical results. A superconductivity enhancement is reported in amorphous GeH. It is revealed that the evolution of $T_c$ arises from the pressure-induced structural transition and phonon softening related to an in-plane Ge-Ge phonon mode. The amorphous GeH superconductor offers a platform for studying amorphous hydride superconducting phenomena.

## 4. Experimental Section

*Sample preparation*: In this paper, the germanane (GeH) crystals were prepared using the most popular topological deintercalation method, as described in Ref. 11. In this method, $CaGe_2$ crystals were first synthesized. High purity calcium (Ca) and germanium (Ge) in a stoichiometric ratio of 1:2 were sealed in an evacuated quartz tube. The quartz tube was quickly heated to and held at 1000 °C in a furnace for 25 hours, then cooled down to room temperature at 25 °C. To synthesize the GeH product, the $CaGe_2$ crystals were completely immersed in concentrated HCl acid for a week at -40 °C. This reaction is a displacement process, in which the Ca atoms in $CaGe_2$ are replaced by H atoms in concentrated HCl acid to produce the GeH structure. After the reaction, the GeH was washed with deionized water and anhydrous methanol followed by deionized water to remove contaminants.

*Sample characterization at ambient pressure*: Powder X-ray diffraction (XRD, PANalytical X9 PertPro X-ray diffractometer using Cu Kα radiation) was performed to study the structure of GeH. The surface morphologies of the GeH samples were investigated using scanning electron microscopy (SEM, JEOL JSM-7500FA). Room-temperature Raman measurements were performed on an In Via Reflex Raman spectrometer with a laser at 532 nm. The details of the crystal structure were further detected by transmission electron microscopy (TEM, JEM-2011F, JEOL operating at 200 kV). X-ray photoelectron spectroscopy (XPS) was performed on a Thermo Escalab 250XI photoelectron spectrometer using monochromatic Al Kα radiation under vacuum at $1 \times 10^{-10}$ mbar. Fourier transform infrared (FTIR) spectra were collected on a Shimadzu FTIR Prestige-21 using KBr as the reference sample. Elemental analysis was performed on an energy dispersive X-ray spectroscopy (EDS, OXFORD INCA X-ACT).

*In-situ high-pressure synchrotron XRD at room temperature*: All high-pressure experiments were carried out by using a symmetric DAC with a 400 mm diameter culet. The GeH sample and a ruby ball were loaded inside a sample chamber, which was laser-drilled to a diameter of 140 μm in a T301 steel gasket that was pre-indented to a thickness of 50 μm. Silicone oil served as the PTM. High-pressure XRD measurements were carried out at the High-Pressure Station of the Beijing Synchrotron Radiation Facility. The wavelength of the X-ray beam was 0.6199 Å. $CeO_2$ was used as a standard sample for calibration. The 2D XRD images were integrated to one-dimensional (1D) patterns with Dioptas software.

*In-situ high-pressure Raman spectroscopy*: High-pressure Raman at room temperature are performed on a Renishaw inVia Raman Microscope with an $Ar^+$ 514.5 nm laser, 1800 g/mm grating and 500 nm blazed wavelength. High pressure was generated by a DAC without PTM. The pressures were determined from the pressure dependent shift of the R1 line fluorescence of ruby. Before high-pressure Raman measurement, the GeH sample is measured from low intensity of Raman peak to high intensity until the sample is damage. We choose the laser power before damage for high pressure Raman measurement. In this work, GeH starts to damage when the power is 1.25 mW, thus we keep the laser power at 0.80 mW. We have also carried out *in-situ* high-pressure Raman spectroscopy with silicone oil served as the PTM. We have also performed *in-situ* high-pressure Raman with silicone oil as PTM at 25 K.

*High-pressure resistivity*: Electrical resistance under high pressure was measured by using a four-probe resistance test system in a DAC at pressures up to 34.11 GPa without PTM. Resistance was measured using an inverting *dc* current in a van der Pauw technique implemented in a customary cryogenic setup (lowest achievable temperature $T_{min}$= 2.5 K). Firstly, a 250 mm thick metallic gasket of Re was indented with about 17–20 GPa pressure.

Then, the bottom of the imprint of diameter about 400 μm was drilled out. An insulating gasket is required to separate the metallic gasket from the electrodes. $Al_2O_3$ powder and epoxy mixture was used for the insulating Re gaskets, while gold foil was performed in the electrical leads. The diameters of the flat working surface of the diamond anvil and the sample chamber were 400 μm. The samples and a tiny ruby ball beside to calibrate pressure were loaded into the DAC. The resistance values were defined as an average of five successive measurements at constant temperature. The magentoelectric meausrement is similar to the previous studies.[48-49] The data were collected in a DAC made of nonmagnetic Cu–Be alloy. The diamond culet was 300 mm in diameter. The DAC was placed inside a homemade multifunctional measurement system (1.8–300 K, JANIS Research Company Inc.; 0–9 T, Cryomagnetics Inc.). Helium (He) as the medium for heat convection to precise temperature control and obtain a high efficiency of heat transfer. Two Cernox resistors (CX-1050-CU-HT-1.4L) located near the DAC were used to ensure the accuracy of the temperature in the presence of a magnetic field.

*DFT calculations*: A stable crystal structure was predicted for GeH at 10 GPa using the particle-swarm optimization methodology based Crystal Structure Analysis software package with Particle Swarm Optimization (CALYPSO) code.[50,51] Optimization of the electronic structures of GeH was performed using density functional theory (DFT) within the Perdew-Burke-Ernzerhof parameterization of generalized gradient analysis (GGA), as implemented in the Vienna ab initio simulation package (VASP) code.[52] The projector augmented wave (PAW) method was employed to describe electron-ion interactions.[53] Brillouin zone (BZ) sampling used a grid with spacing of 15×15×11 and a plane-wave basis cut-off set at 500 eV. Equilibrium geometries were obtained by the minimum energy principle until the energy and force converged to $10^{-6}$ eV and $10^{-4}$ eV/Å, respectively. Phonon dispersion curves and normal modes were obtained by applying the supercell method in PHONOPY code.[54] Lattice dynamics and electron-phonon coupling (EPC) calculations of GeH at 10 GPa were performed with the QUANTUM-ESPRESSO package[55] using Optimized Norm-Conserving Vanderbilt (ONCV) pseudopotentials[56] and plane wave basis sets with a kinetic energy cut-off of 180 Ry. A 24 × 24 × 24 Monkhorst-Pack k-point grid with a Gaussian smearing of 0.01 Ry and a 2 × 2 × 2 q-point mesh in the first Brillouin zone was used for the calculation of EPC matrix elements.

**Acknowledgements**

This work was financially supported by Beijing Natural Science Foundation (Z180007), the National Key R&D Program of China (2018YFE0202700 and 2018YFA0305900), Australian Research Council (DP170101467, FT180100585 and LP180100722), National Natural Science Foundation of China (52073006, 51672018, 51472016, 11874003, 11874172, U2032215, 11904015, 12074021, 51320105007). The authors acknowledge the supports from the UOW-BUAA Joint Research Centre. The authors thank Dr. H. F. Feng and Dr. T. Silver for proof reading and valuable discussion.

Received: ((will be filled in by the editorial staff))
Revised: ((will be filled in by the editorial staff))
Published online: ((will be filled in by the editorial staff))

Sclauzero, A. P Seitsonen, A. Smogunov, P. Umari, R. M Wentzcovitch, *J. Phys. Condens. Matter* **2009**, 21, 395502.

[56] D. R. Hamann, *Phys. Rev. B* **2013**, 88, 085117.

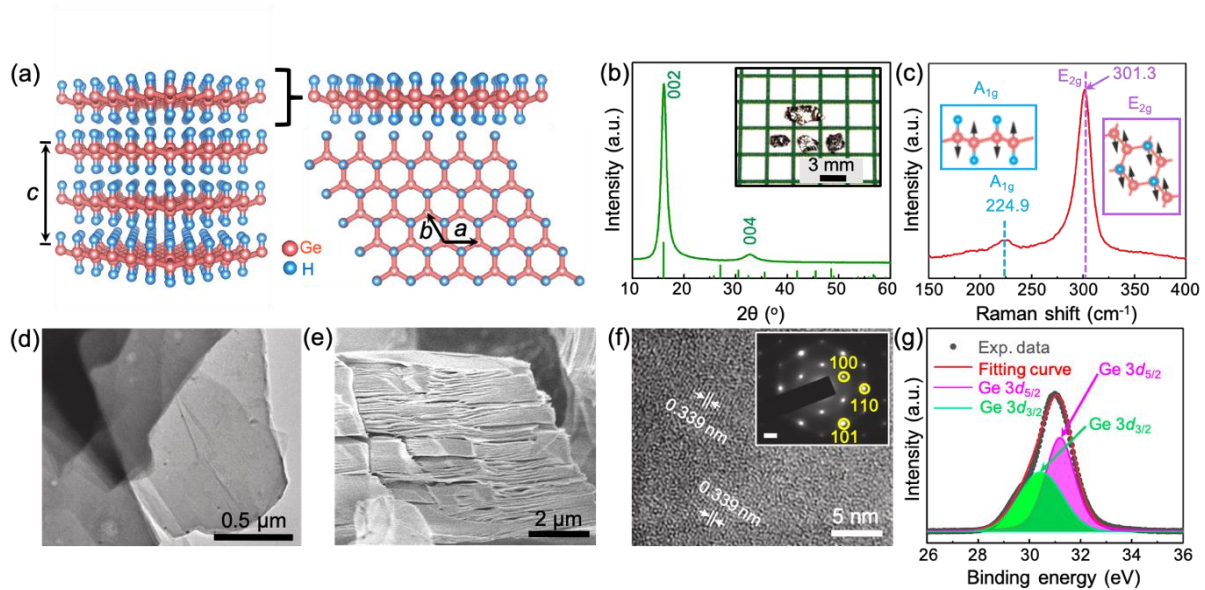

**Figure 1.** Structural characterization of 2H GeH. **a** The crystal structure of the vdW layered GeH. The red and blue balls represent the Ge and H atoms, respectively. **b** Powder XRD pattern of GeH single crystal (inset: optical image of GeH single crystal). **c** Raman spectra of GeH product. The insets show phonon modes of GeH, where $E_{2g}$ represents Ge-Ge in-plane vibration and $A_{1g}$ represents GeH out-of-plane vibration. **d** TEM image of GeH flake, and **f** high-resolution TEM image of GeH. Inset is the corresponding SAED pattern of GeH, which demonstrates its honeycomb structure. Scale bar is 21/nm. **e** SEM image of GeH indicates its layered structure. **g** The XPS spectrum of Ge 3*d* in the GeH sample.

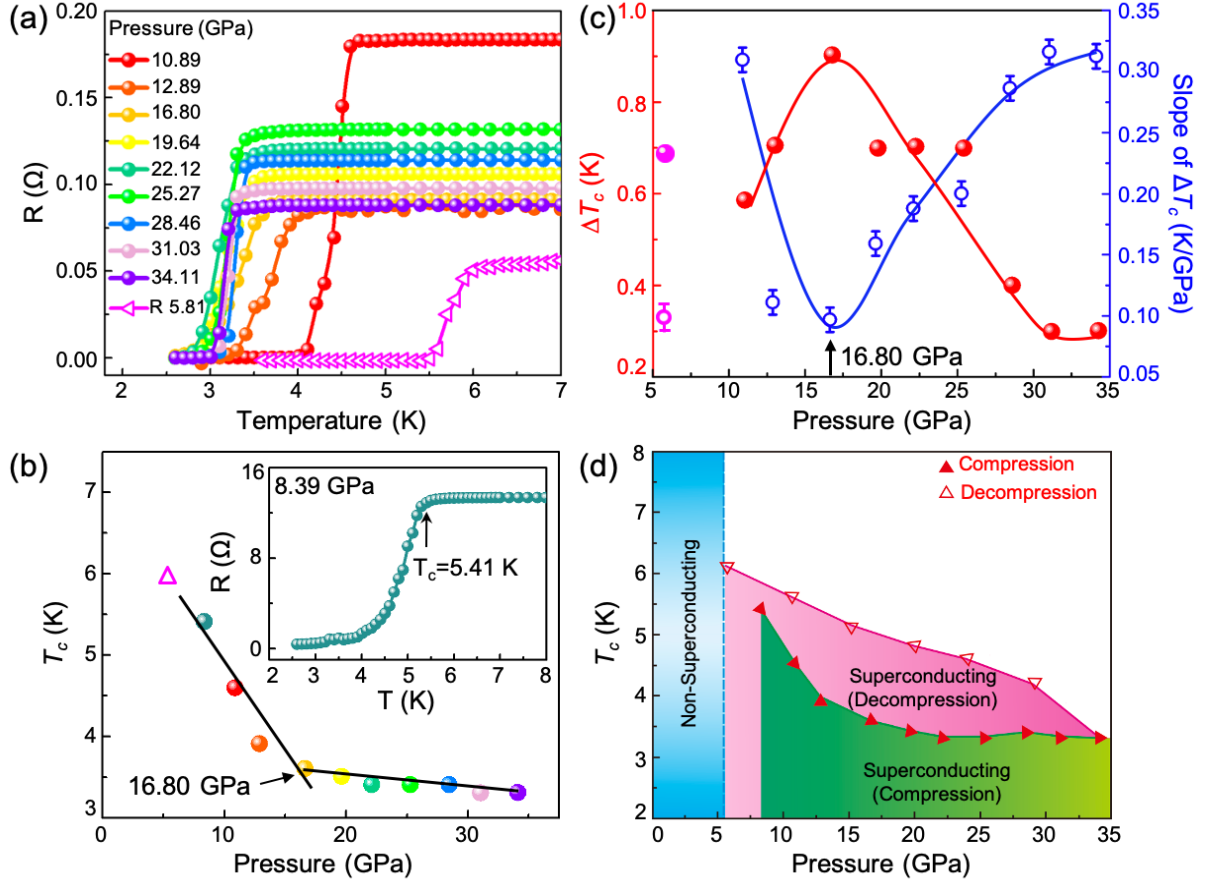

**Figure 2. a** Temperature-dependence of the electrical resistance at representative pressures during compression and decompression using a DAC under zero magnetic field. **b** Superconducting transition temperature as a function of pressure of the GeH sample. The inset is the temperature dependence of the electrical resistivity of GeH at 8.39 GPa. **c** The pressure-dependent superconducting transition width and the slope of superconducting transition width. **d** Temperature-pressure phase diagram of GeH from resistance measurements. The solid and open triangle symbols represent the superconducting transition temperature values during compression and decompression, respectively.

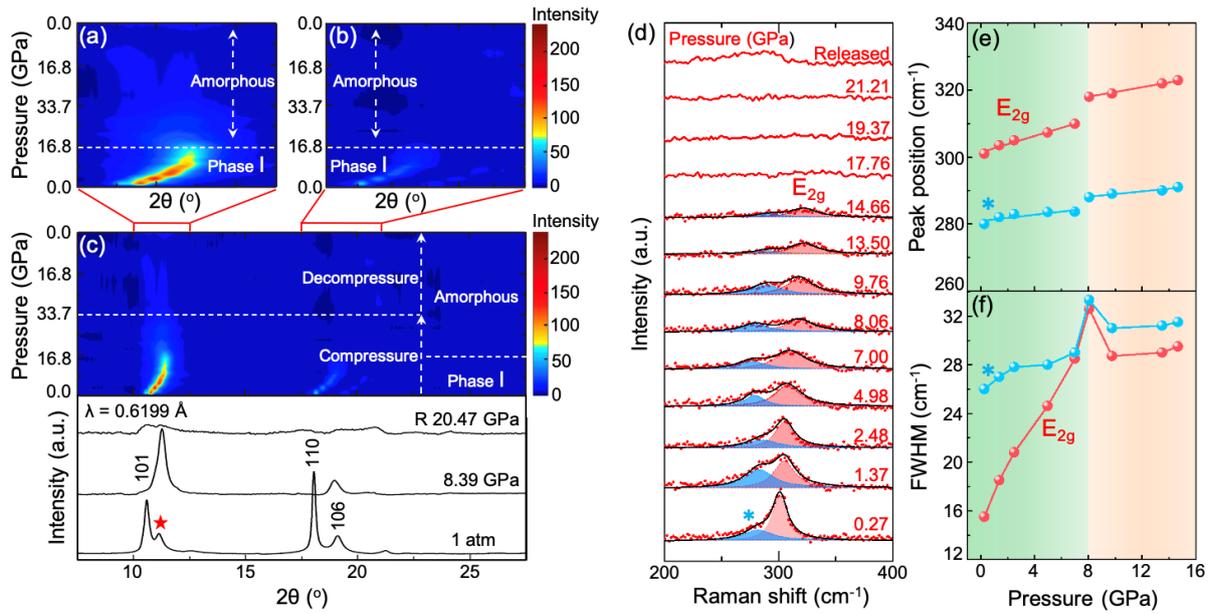

**Figure 3.** *In-situ* high-pressure synchrotron XRD and *in-situ* high-pressure Raman spectroscopy results. **c** Intensity contour map of the synchrotron XRD patterns of GeH collected during compression up to 33.71 GPa and decompression to ambient conditions at room temperature (phases identified in **a** and **b**). The incident wavelength was $\lambda = 0.6199$ Å. **d** Raman spectra of GeH sample measured at different pressures. All spectra were collected at room temperature. The Raman peak marked with blue asterisk is assigned to the Raman splitting caused by layer sliding under high pressure. **e** Pressure dependence of the Raman peak position for the sample. **f** Pressure dependence of the Raman FWHM for the sample.

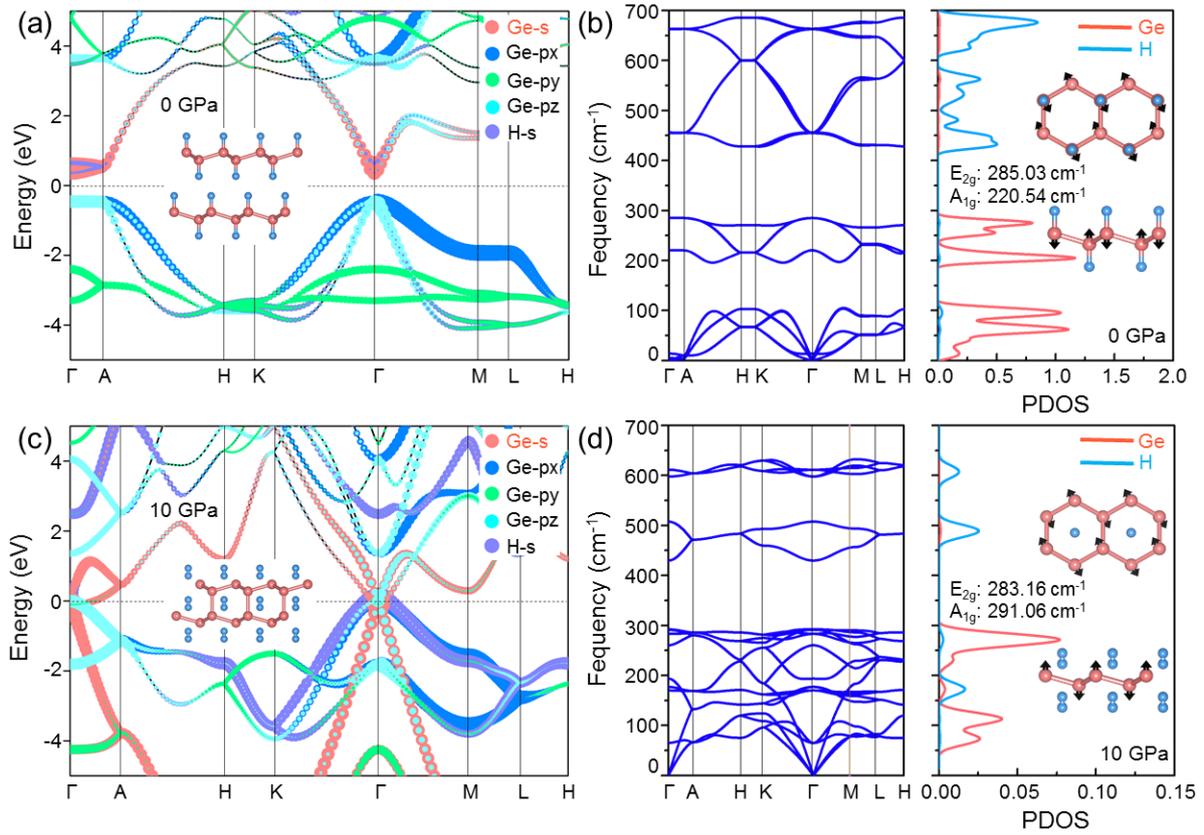

**Figure 4.** Electronic and superconducting properties of 2H GeH. **a** Partial band structure projection of atoms and crystal structure at 0 GPa. **b** Phonon dispersion and phonon density of states at 0 GPa. **c** Partial band structure projection of atoms and crystal structure at 10 GPa. **d** Phonon dispersion and phonon density of states at 10 GPa.

22